\documentclass[aps,prd,reprint,superscriptaddress,showpacs,showkeys]{revtex4-2}

\usepackage{graphicx}
\usepackage{dcolumn}
\usepackage{bm}
\usepackage{natbib}
\usepackage{color}
\usepackage{subfigure}
\usepackage{array}
\usepackage{makecell}
\usepackage{url}
\usepackage{hyperref}
\usepackage{multirow}

\begin{document}

\title{Measurement of the Compton scattering in germanium with a p-type point-contact germanium detector for dark matter detection}

\author{J.~W.~Hu}
\affiliation{Key Laboratory of Particle and Radiation Imaging (Ministry of Education) and Department of Engineering Physics, Tsinghua University, Beijing 100084}

\author{L.~T.~Yang}
\email{Corresponding author: yanglt@mail.tsinghua.edu.cn}
\affiliation{Key Laboratory of Particle and Radiation Imaging (Ministry of Education) and Department of Engineering Physics, Tsinghua University, Beijing 100084}

\author{Q.~Yue}
\email{Corresponding author: yueq@mail.tsinghua.edu.cn}
\affiliation{Key Laboratory of Particle and Radiation Imaging (Ministry of Education) and Department of Engineering Physics, Tsinghua University, Beijing 100084}

\author{X.~P.~Geng}
\affiliation{Key Laboratory of Particle and Radiation Imaging (Ministry of Education) and Department of Engineering Physics, Tsinghua University, Beijing 100084}

\author{H.~B.~Li}
\affiliation{Institute of Physics, Academia Sinica, Taipei 11529}

\author{Y.~F.~Liang}
\affiliation{Key Laboratory of Particle and Radiation Imaging (Ministry of Education) and Department of Engineering Physics, Tsinghua University, Beijing 100084}

\author{S.~T.~Lin}
\affiliation{College of Physics, Sichuan University, Chengdu 610065}

\author{S.~K.~Liu}
\affiliation{College of Physics, Sichuan University, Chengdu 610065}

\author{H.~Ma}
\affiliation{Key Laboratory of Particle and Radiation Imaging (Ministry of Education) and Department of Engineering Physics, Tsinghua University, Beijing 100084}

\author{L.~Wang}
\affiliation{Department of  Physics, Beijing Normal University, Beijing 100875}

\author{H.~T.~Wong}
\affiliation{Institute of Physics, Academia Sinica, Taipei 11529}

\author{B.~T.~Zhang}
\affiliation{Key Laboratory of Particle and Radiation Imaging (Ministry of Education) and Department of Engineering Physics, Tsinghua University, Beijing 100084}

\date{\today}

\begin{abstract}
Low-energy background through Compton scattering from the ambient $\gamma$ rays can be contaminated in direct dark matter search experiments. In this paper, we report comparable measurements on low-energy spectra via Compton scattering from several $\gamma$-ray sources with a p-type point-contact germanium detector. The spectra between 500 eV and 18 keV have been measured and analyzed. Moreover, the features of the electron binding effect, particularly at the edges of the K- and L-shells in the germanium atom, were observed with different gamma sources and were consistent with the models in the Geant4 simulation. An empirical background model is proposed that provides insights into understanding the low-energy background in germanium for direct dark matter experiments.

\end{abstract}

\maketitle

\section{Introduction}\label{section1}
Direct dark matter searches can be performed by measuring the energy deposited by the scattering of an incident dark matter particle with a nucleon or an electron in the target. For dark matter detection experiments, one of the most important and challenging issues is to further decrease the background level so as to achieve higher experimental sensitivity. The small-angle Compton scattering of ambient $\gamma$ rays is one of the main backgrounds in dark matter detection experiments. Therefore, a better understanding of the Compton scattering process of the ambient $\gamma$ rays on dark matter experimental detectors can allow more accurate energy spectral analyses in low-energy regions and further improve the sensitivity of direct dark matter experiments.

Ramanathan $et\ al.$ measured Compton spectra between 60 eV and 4 keV in silicon using silicon charge-coupled devices, and they first observed and reported the steps of Compton scattering for silicon detectors~\cite{DAMIC-PRD2017}. Recently, more silicon measurements are reported, which go to lower energy (23 eV), single electron resolution allowing the features associated with the silicon $\rm L_1$- and $\rm L_{2,3}$-shells to be clearly differentiated, and better precision to understand the shape of the spectrum~\cite{DNorcini2022, AMBotti2022}. The SuperCDMS collaboration compared the calibration data in SuperCDMS-Soudan experiments using \texttt{Geant4.10} simulations~\cite{CDMSlite:2016eil} and observed the Compton steps at an energy of $\sim$11.1 keV associated with the K-shell structure of a germanium atom. The p-type point contact germanium (pPCGe) detector used in the China Dark Matter Experiment (CDEX)~\cite{cdex12014, cdex12016, cdex1b2018, cdex_migdal, cdex_am, cdex_axion, cdex102018, c10_darkphoton} on light dark matter searches could reach an ultralow energy threshold of 160 eV and has good energy resolution~\cite{cdex1b2018, cdex_migdal, cdex_am}, thereby facilitating spectra measurement at their energy level, which is $\sim$1.2 keV associated with the L-shell binding energy of germanium atoms.

In this study, using the CDEX-1B pPCGe detector, we investigated the shapes and fine structures of the energy spectra in the low-energy region produced by the $\gamma$ rays of external radioactive sources. In these experimental energy spectra, we measured the L-shell step for the first time, as well as the K-shell step at the low-energy region of the germanium detector~\cite{CDMSlite:2016eil,SuperCDMS-PhysRevD2019}. Moreover, using the \texttt{Livermore} Compton model in Geant4, we simulated the low-energy region spectra induced by external $\gamma$ rays in the HPGe detector. The obtained experimental measurements and simulated spectra were in good agreement.

\section{Compton Scattering}\label{sec:2}
Compton scattering is an electromagnetic process where an incident photon ($\gamma$) transfers some of its energy to an electron, resulting in a recoiled electron and a scattered photon deflected from its original direction~\cite{DAMIC-PRD2017,Compton-PhysRev1923}. For an interaction with a free electron at rest, the differential cross section, which is related to the scattering angle ($\theta=0-\pi$), can be given by the well-known Klein-Nishina formula~\cite{Klein1929}. With the continuous change of scattering angle, a continuous energy transferred to the electron ($E=E_{\gamma}-E_{s}$) can be obtained from zero up to the Compton edge, which is a spectral feature stands for the maximum energy deposited in the interaction and occurs when the $\gamma$ ray backscatters, i.e., when $\theta=\pi$. The energy of the scattered photon $E_{s}$ depends on the incident photon energy $E_{\gamma}$, electron mass $m_{e}$, and scattering angle $\theta$, as~\cite{DAMIC-PRD2017,AMBotti2022}
\begin{equation}\label{eq_1}
E_{s}=\frac{E_{\gamma}}{1+\frac{E_{\gamma}}{m_{e}c^{2}}\left(1-cos\theta\right)}.
\end{equation}

For bound electrons, the electrons in atomic shells have a certain momentum distribution with discrete energy levels which must be taken into account when calculating the cross section. According to the relativistic impulse approximation (RIA)~\cite{Ribberfors-PhysRevB1975,Ribberfors-PhysRevA1982,Ribberfors-PhysRevA1984,PHolm-PhysRevA1988}, the cross section has jumps near the shell electron binding energy. A summary of the germanium atomic shells is given in Table~\ref{tab:tab1}. The RIA formulation can be obtained by modifying the KN formula as follows~\cite{Ribberfors-PhysRevA1982}
\begin{equation}\label{eq_2}
\frac{d^{2}\sigma}{dE_{s}d\Omega}\simeq\frac{r_{0}^{2}m_{e}^{2}E_{s}}{2E_{\gamma}|\overrightarrow{k_{1}}-\overrightarrow{k_{2}}|\left(m_{e}^{2}+p_{z}^{2}\right)^{2}}\times\overline{X}\left(R_{1}, R_{2}\right)\times J\left(p_{z}\right),
\end{equation}
where $\frac{d^{2}\sigma}{dE_{s}d\Omega}$ is the double-differential Compton scattering cross section (DDCS), $r_{0}$ is the classical radius of electron, $|\overrightarrow{k_{1}}-\overrightarrow{k_{2}}|$ is the modulus of the momentum transfer vector, $p_{z}$ is the electron's initial momentum on the momentum transfer direction, and $\overline{X}\left(R_{1}, R_{2}\right)$ is a slowing varying function~\cite{Ribberfors-PhysRevA1982}. When relativistic effects are less significant, such as relatively low-energy and momentum transfers, the function $\overline{X}\left(R_{1}, R_{2}\right)$ can be simplified to a KN-type expression. $J\left(p_{z}\right)$ are the Compton profiles, which encode the momentum distribution of target electrons before the collision. The total cross section can be evaluated by numerically integrating Eq.~\ref{eq_2}.

\begin{table}[!htbp]
\begin{ruledtabular}
\caption{Germanium atomic shells. The table details the quantum numbers ($n,l$), binding energy ($E_{nl}$), and number of electrons in each atomic shell. The binding energies are from experimental x-ray measurements at National Institute of Standards and Technology~\cite{NIST_web,XrayDataBooklet_2009}.}
\label{tab:tab1}
\centering
\begin{tabular}{ccccc}
Shell & $n$ & $l$ & Energy (eV) & Electrons\\
\hline
K& 1 & 0 & 11103 & 2 \\
\hline
$\rm L_1$ & 2 & 0 & 1414 & 2\\
\hline
$\rm L_2,L_3$ & 2 & 1 & 1248, 1217 & 6\\
\hline
$\rm M_1$ & 3 & 0 & 180 & 2\\
\hline
$\rm M_2,M_3$ & 3 & 1 & 125, 121 & 6\\
\hline
$\rm M_{4,5}$ & 3 & 2 & 29 & 10\\
\hline
Valence & 4 & ... & 0.72 & 4\\
\end{tabular}
\end{ruledtabular}
\end{table}

In the low-energy transfer region, the energy spectrum from Compton scattering approximately presents multiple linear platforms and step structures due to the differences in the number of electrons and average binding energies in atomic shells. This linearity is particularly evident when an energy transfer $E$ approaches the photoionization thresholds, which are 11.1 keV for the K-shell electrons and 1.4 keV for the L-shell electrons of a Ge atom. This enables the use of linear functions to describe energy spectra in the near photoionization threshold region. Thus,
\begin{equation}\label{eq_3}
\frac{d\sigma}{dE}\approx aE+b.
\end{equation}

Under the free electron approximation, the step height ratio is the ratio of the number of electrons from the corresponding shells. However, due to the different Compton scattering cross sections of electrons in each layer, some differences occur. In RIA calculations, for various subshells, their corresponding platforms and slopes, $a_{K}$, $a_{L1}$, $a_{L2}$, $a_{M1}$, can be obtained by the linear fitting of Eq.~\ref{eq_3}, respectively~\cite{Qiao_2020}. The slope of the spectrum between the Compton steps can be approximate estimated with the KN formula, which can reduced to $d\sigma/dE=1-(mc^2/E_{\gamma}^2)E$ when $E/E_{\gamma}\ll 1$~\cite{DAMIC-PRD2017}. For the energies of interest ($E < 20$ keV), the slope approximates a constant when $E_{\gamma}$ is large.

\section{Measurement and data analysis}\label{sec:3}
The CDEX-1B experiment, located in the China Jinping Underground Laboratory with about 2400 m of rock overburden~\cite{cjpl}, utilizes a 1-kg pPCGe detector for light dark matter searches~\cite{cdex1b2018,cdex_migdal,cdex_am,cdex_axion}. The germanium crystal has a height of 62.3 mm, a diameter of 62.1 mm, and a mass of 1008 g. The vacuum cryostat is made of copper; from inside to outside, the germanium crystal is surrounded by an 0.5-mm-thick polytetrafluoroethylene (PTFE) foil and then fixed to a copper cup with a thickness of 2.0 mm (side) and 1.0 mm (top), respectively. The energy threshold of CDEX-1B reached 160 eV, and the dead layer was measured to be 0.88$ \pm $0.12 mm~\cite{cdex1b_deadlayer}. The output signal of the $p^{+}$ point electrode of the germanium crystal was obtained by a pulse optical feedback charge-sensitive preamplifier.

\begin{figure}[!htbp]
\centering\includegraphics[width=\columnwidth]{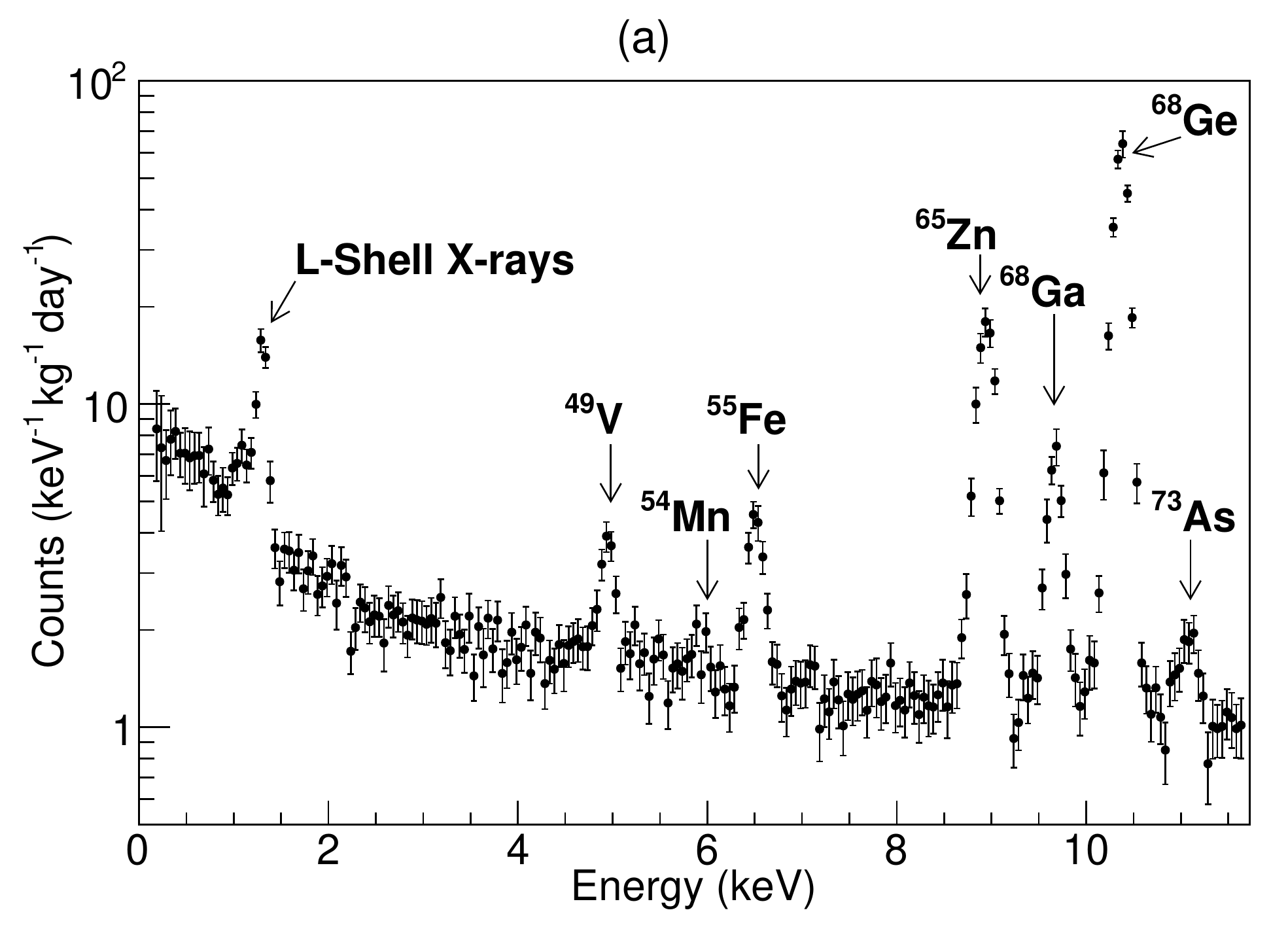}
\centering\includegraphics[width=\columnwidth]{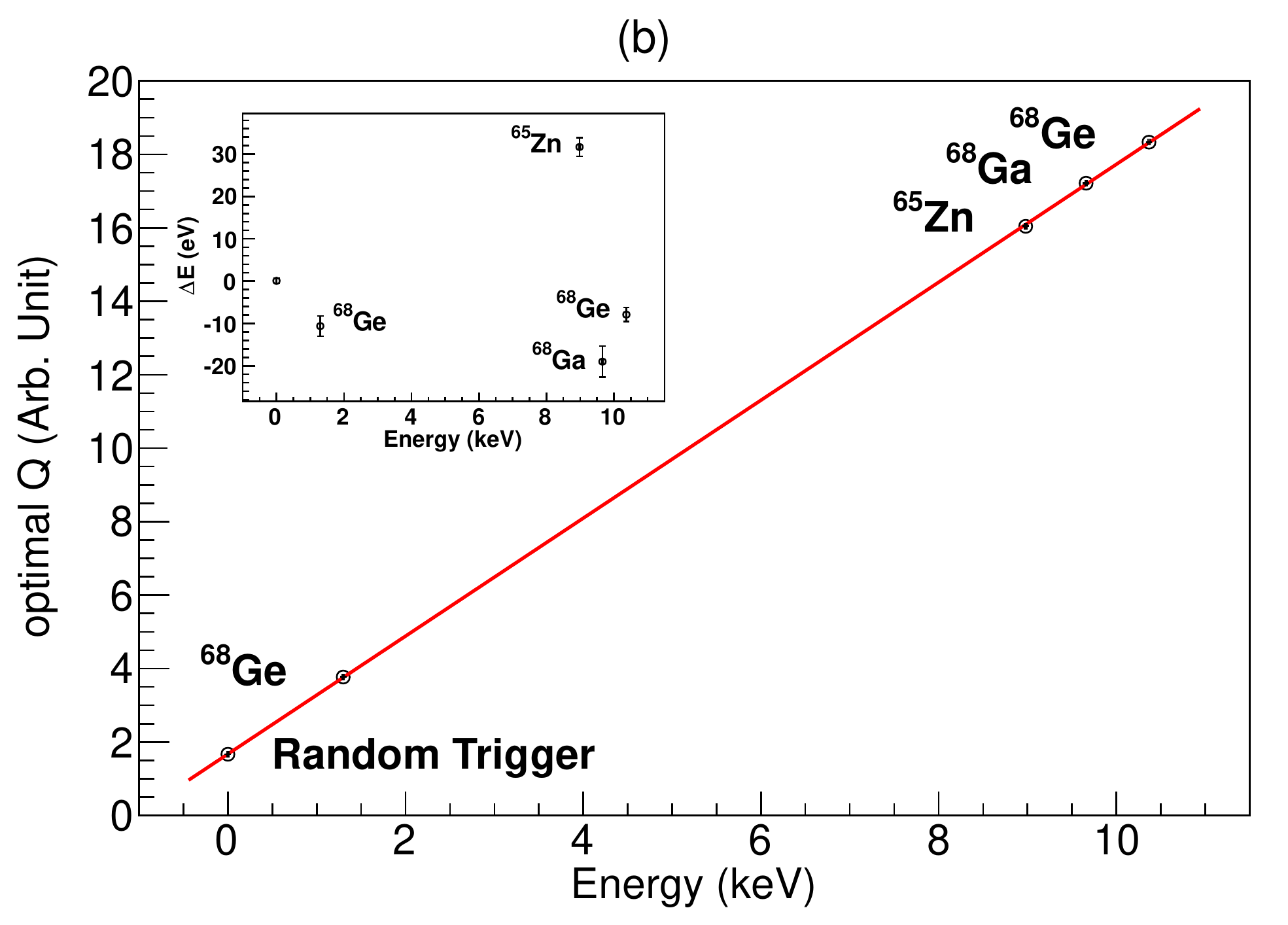}
\centering\includegraphics[width=\columnwidth]{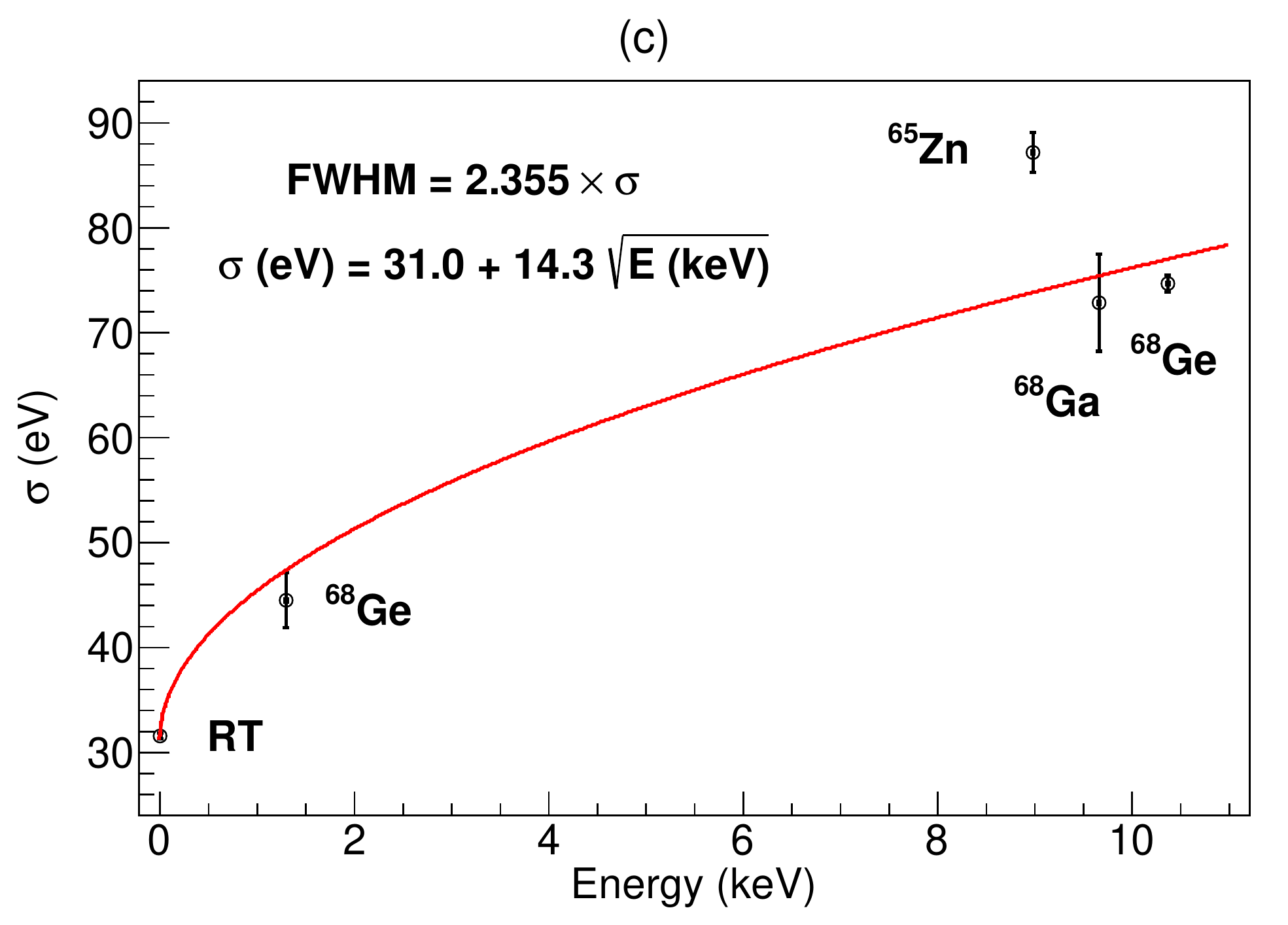}
\caption{(a) Measured energy spectra of CDEX-1B used in previous dark matter analysis~\cite{cdex1b2018,cdex_axion}, these characteristic K-shell x-ray peaks from internal cosmogenic radionuclides are used for energy calibration; (b) energy calibration at the low-energy region with inset showing the energy deviation; (c) energy resolution of CDEX-1B detector.}
\label{fig:1}
\end{figure}

The main performance parameters of CDEX-1B are measured and depicted in Fig.~\ref{fig:1}~\cite{cdex1b2018,cdex_migdal,cdex_am,cdex_axion}. The pedestal RMS achieves 31 eV and the full width at half maximum (FWHM) of a pulser input is 80 eV. The FWHM at 10.37 keV x-rays peak from $^{68,71}$Ge is (177±3) eV. With all these improvements, a physics analysis threshold of 160 eV is achieved, which is significantly lower than earlier achievement by CDEX-1~\cite{cdex12014,cdex12016}. The energy calibration at the low-energy region ($<$20 keV) was achieved using the following internal cosmogenic Xray peaks: $^{68,71}$Ge (KX-line 10.37 keV and LX-line 1.298 keV), $^{68}$Ga (KX-line 9.66 keV), and $^{65}$Zn (KX-line 8.98 keV and LX-line 1.094 keV). The zero energy was defined by the random trigger events (RT, 0 keV). The linearity was so good that the deviation was less than 0.4\%~\cite{cdex1b2018}.

\begin{figure}[!tbp]
\centering\includegraphics[width=\columnwidth]{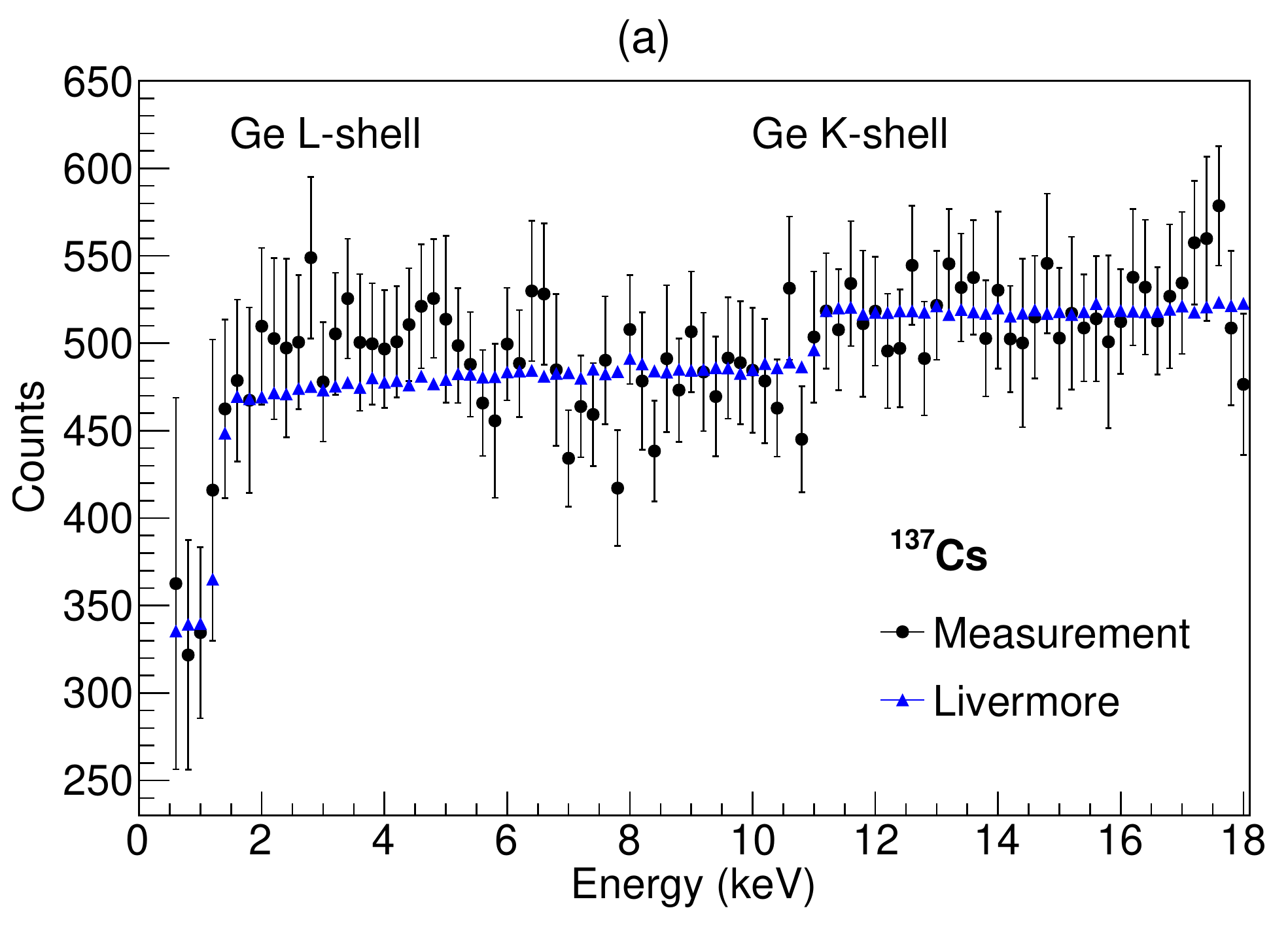}
\centering\includegraphics[width=\columnwidth]{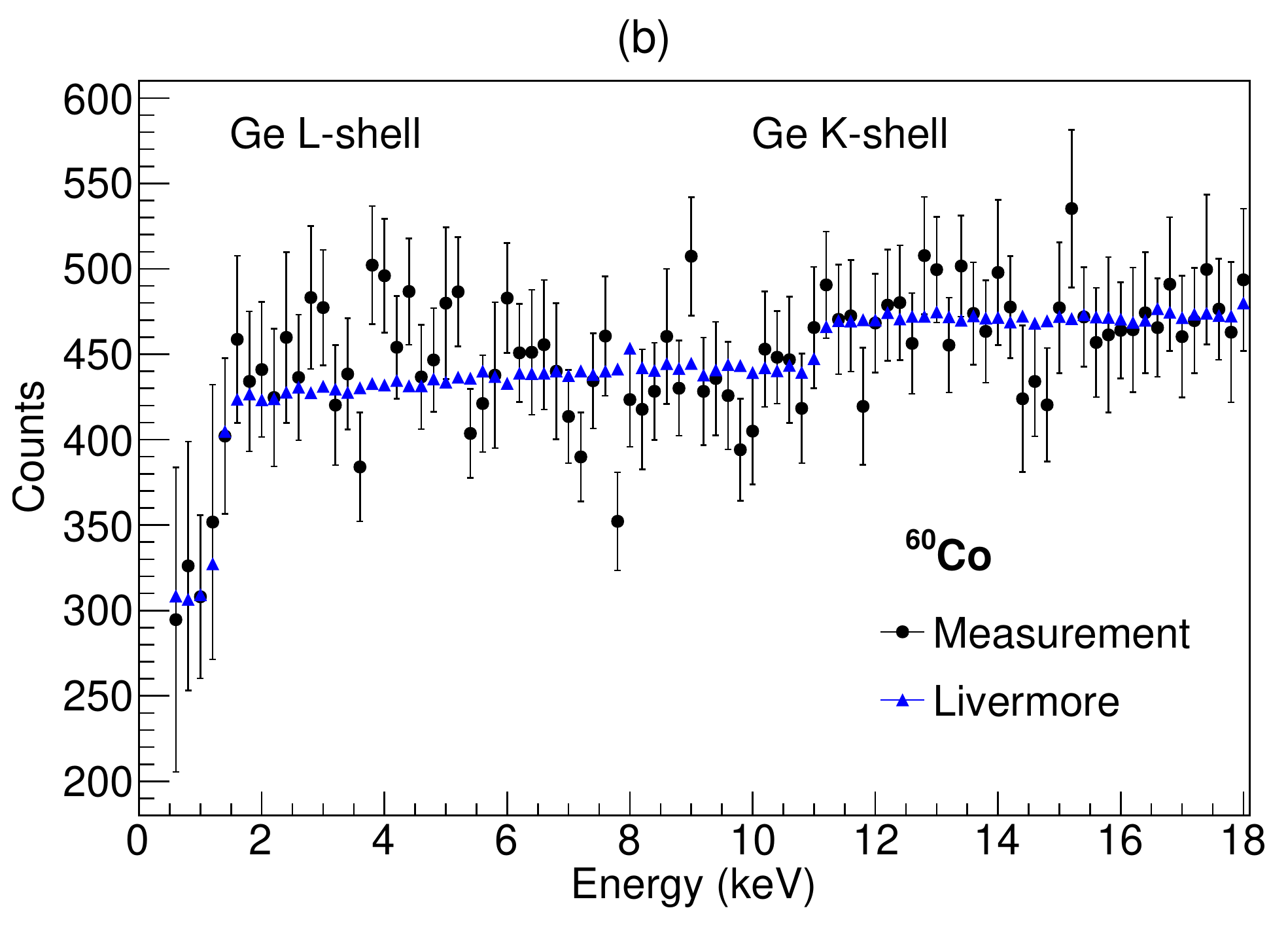}
\caption{Energy spectra of $^{137}$Cs (a) and $^{60}$Co (b). The bin width is 200 eV and the energy range is 0.5$-$18 keV. The simulated results using \texttt{Livermore} model from Geant4 were superimposed for comparison.}
\label{fig:2}
\end{figure}

Two radioactive sources, including $^{137}$Cs and $^{60}$Co, were used for the experimental measurements, and they were placed at the top position of the detector at a distance of 22.2 cm from the copper cryostat. Following the same procedures of physics event selection, associated efficiency correction and systematic uncertainties, as described in detail in our previous work~\cite{cdex1b2018,Yang:2018a}, we constructed spectra of the real bulk events of each calibration source. We focused on the spectra of the CDEX-1B pPCGe detector at the low-energy region below 20 keV. Two obvious steps of the L-shell and K-shell of the germanium nuclei around 1.4 keV and 11.1 keV can be observed in Fig.~\ref{fig:2}. Thus, for the first time, we observed the L-shell step structure in a germanium detector. 

One linear function and two error functions were used to fit the measured Compton steps:
\begin{equation}\label{eq_4}
f(E)=b+\sum_{i=K, L}\frac{h_{i}}{2}\left[1+{\rm erf}\left(\frac{E-E_{i}}{\sqrt{2}\sigma_{i}}\right)\right],
\end{equation}
where $b$ is the sum of step amplitudes of all the shells except for the K- and L-shells, $h_{i}$ are the step amplitudes of the K- and L-shells, $E_{i}$ are the shell binding energies, and $\sigma_{i}$ are the energy resolutions at the respective energies. The step height mainly depends on the number of electrons in each shell and the electron wave function, especially when measuring the step heights of the L-shell and even the M-shell step height, and the electron wave function has a greater impact. It should be noted that according to the RIA theory, the L-shell step has a sublayer step structure; however, since it is limited by energy resolution, it was only analyzed as one step in this work.

\section{CDEX-1B Simulation}\label{sec:4}
Considering the atomic structure of the atom, atomic deexcitation, and the properties of bound electrons, \texttt{G4LivermoreComptonModel} in the Geant4 package has been specifically designed for the region below a few keV~\cite{AGOSTINELLI2003250,ivanchenko:in2p3-00658779}. The National Institute of Standards and Technology (NIST) data shows that in the energy range of 2$-$10 keV, the difference between the Livermore model and NIST data is less than 5\% ~\cite{1495783}. Since the K-shell step is near 11.1 keV, in this study, the Livermore model was adopted. Based on the Compton scattering model proposed by Ribberfors, the \texttt{G4LivermoreComptonModel} adopts the momentum of the electrons transferred within the two-dimensional plane of the incident and scattered photon. The framework can be extended to the Compton scattering of atomic bound electrons by relativistic pulse approximation~\cite{Ribberfors-PhysRevB1975, DuMond-PhysRev1929}.

We performed the $^{137}$Cs and $^{60}$Co sources to irradiate the CDEX-1B detector model so as to compare the \texttt{G4LivermoreComptonModel} in \texttt{Geant4.10.5} with the experimental results. \texttt{G4LivermoreComptonModel} was simulated with a billion incident $\gamma$ rays, and the deposited energy was recorded, as shown in Fig.~\ref{fig:2}. Then, while taking energy resolution into account, the deposition energy spectra were analyzed and compared with the experimental energy spectra.

\begin{figure}[!htbp]
\centering\includegraphics[width=\columnwidth]{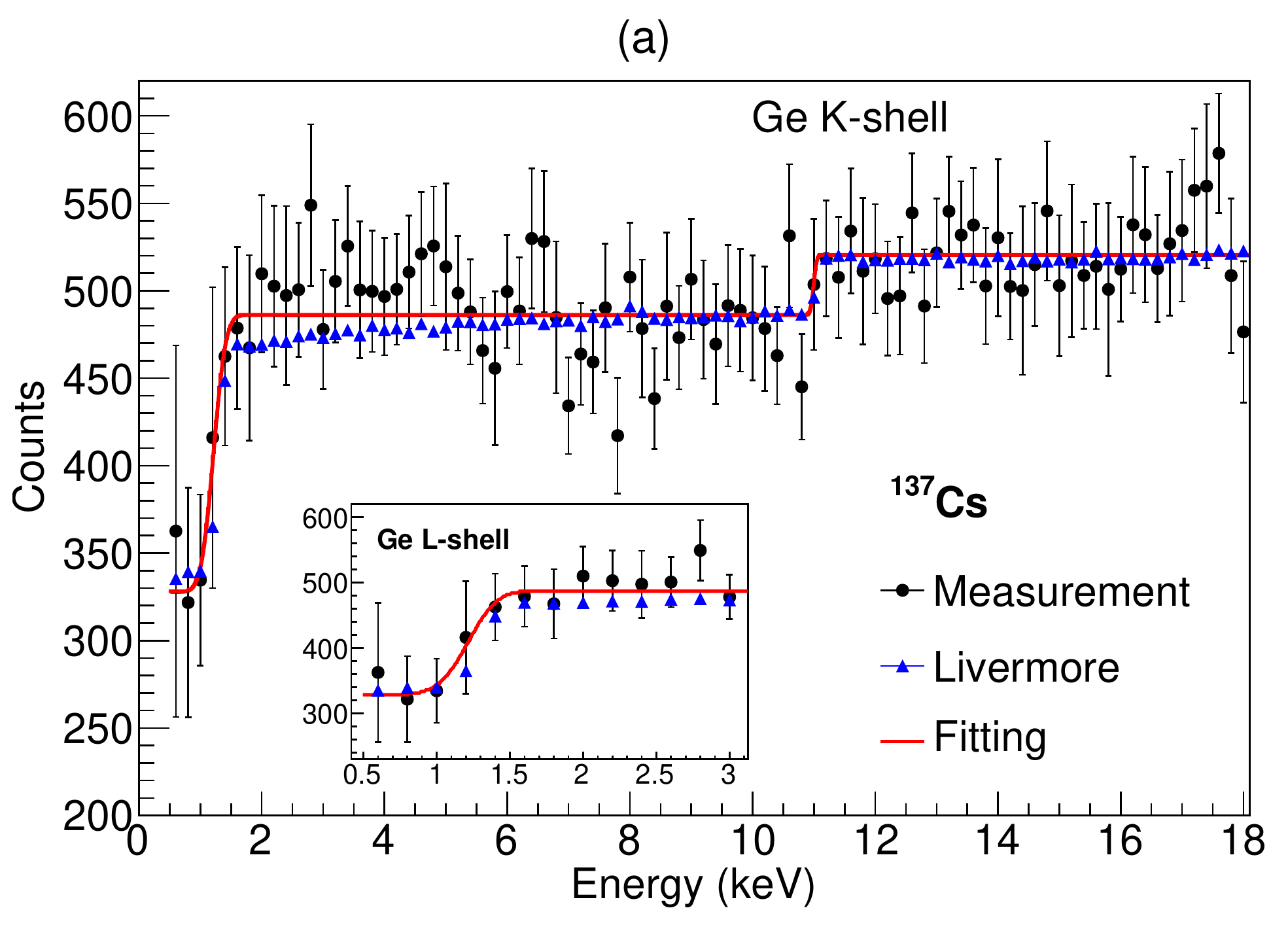}
\centering\includegraphics[width=\columnwidth]{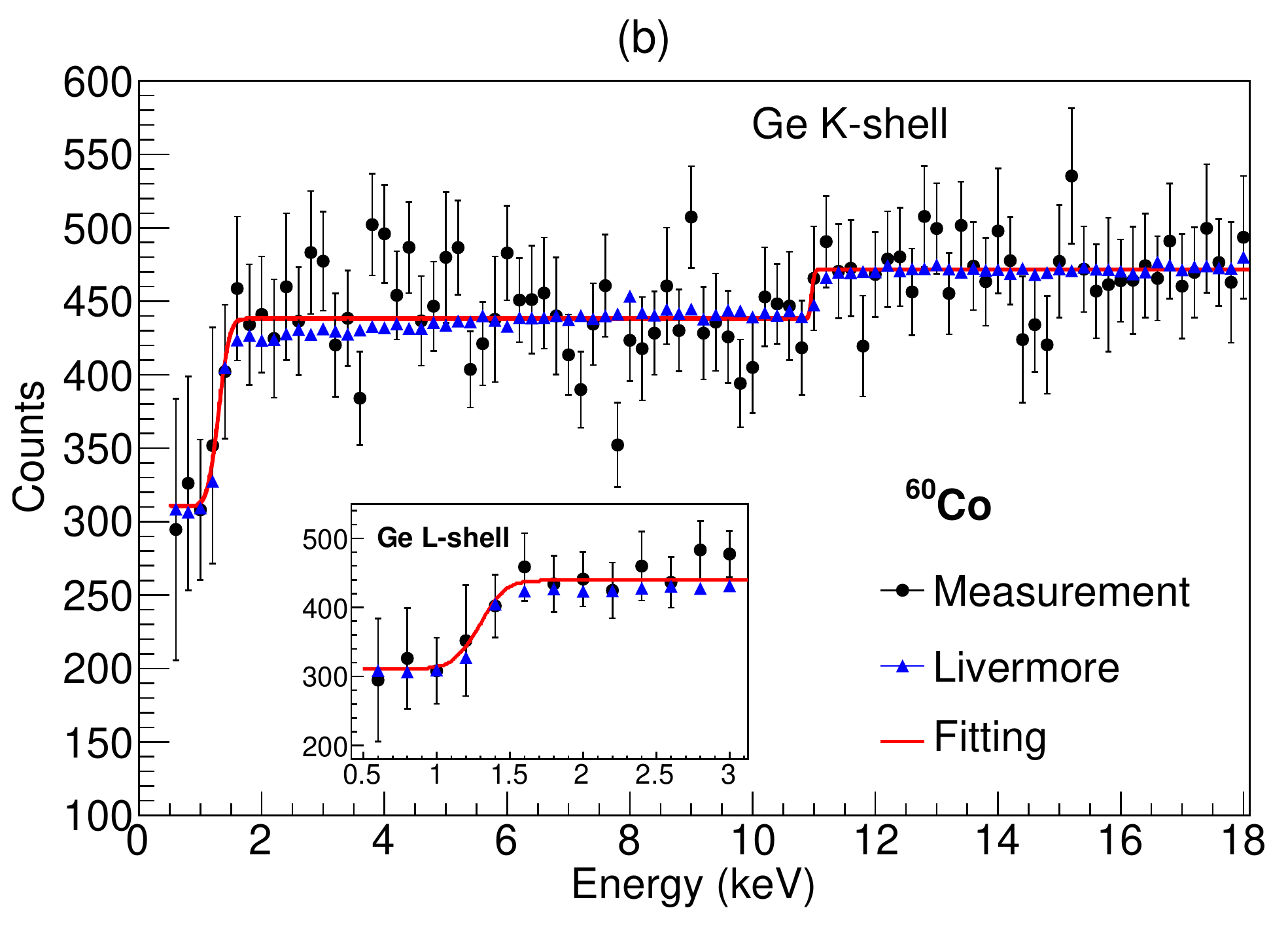}
\caption{Experimental measurement spectra of the $^{137}$Cs (a) and $^{60}$Co (b) at the energy range of 0.5$-$18 keV, with inset showing the newly measured Ge L-shell step.}
\label{fig:3}
\end{figure}

We compared the experimental energy spectra produced by $^{137}$Cs and $^{60}$Co with the energy spectra obtained from the above-mentioned simulations, and the results showed an agreement between the simulated and experimental data in the energy region of 0.5$-$18 keV. However, the uncertainty is dominated by the statistical error, which is still large and limited by the statistical amount of the experimental data in this work. More data will be obtained in the future to reduce the uncertainty and achieve a precision measurement.

To study the structure of the Compton scattering energy spectra in the low-energy region, as measured by the CDEX-1B pPCGe detector, we needed to define the step height ratio $P_{i}$ for each different shell ($i=$K or L):
\begin{equation}\label{eq_6}
P_{i}=h_{i}/(b+h_{K}+h_{L}),
\end{equation}
where $P_{i}$ is the difference between the average heights of two adjacent steps compared with the average height of the step in the upper K shell. Depending on the number of electrons in each shell, the step height ratios for the K- and L-shells should be 6.25\% and 25\%, respectively. The electron wave function affects the Compton step amplitude. Figure~\ref{fig:3} shows the model obtained by fitting the Compton scattering energy spectra due to $\gamma$ rays from $^{137}$Cs and $^{60}$Co using Eq.~\ref{eq_3}. A global fitting from 0.5 keV to 18 keV shows that the fitting function can better describe the experimental measurement data, and then the relative height of the step can be calculated. Table~\ref{tab:tab2} shows the experimental and simulated Compton step results. The experimental errors of $^{137}$Cs and $^{60}$Co include systematic and statistical errors, and the simulation errors of the $^{137}$Cs and $^{60}$Co data are statistical errors. The L-shell step structures are clearly identified in the Compton scattering energy spectra in the low-energy region for both two sources, which is the first time that an L-shell step structure of Ge target atoms in its Compton scattering spectrum is observed at $\sim$1.1 keV. At the same time, we also measured the K-shell step structure at $\sim$11.1 keV~\cite{CDMSlite:2016eil,SuperCDMS-PhysRevD2019}.

\begin{table}[!htbp]
\begin{ruledtabular}
\caption{Compton step parameters for $^{137}$Cs and $^{60}$Co.}
\label{tab:tab2}
\centering
\begin{tabular}{lcc}
Data & $P_{K} [\%]$ & $P_{L} [\%]$  \\
\hline
       $^{137}$Cs Measured & $6.57 \pm 2.71$ & $30.45 \pm 14.64$ \\
\hline
       $^{137}$Cs Simulated & $7.43 \pm 0.10$ & $27.82 \pm 0.24$  \\
\hline
       $^{60}$Co Measured & $7.14 \pm 2.81$ & $27.05 \pm 13.01$ \\
\hline
       $^{60}$Co Simulated & $7.60 \pm 0.14$ & $27.68 \pm 0.32$  \\
\end{tabular}
\end{ruledtabular}
\end{table}

\begin{figure*}[!htbp]
\centering\includegraphics[width=\columnwidth]{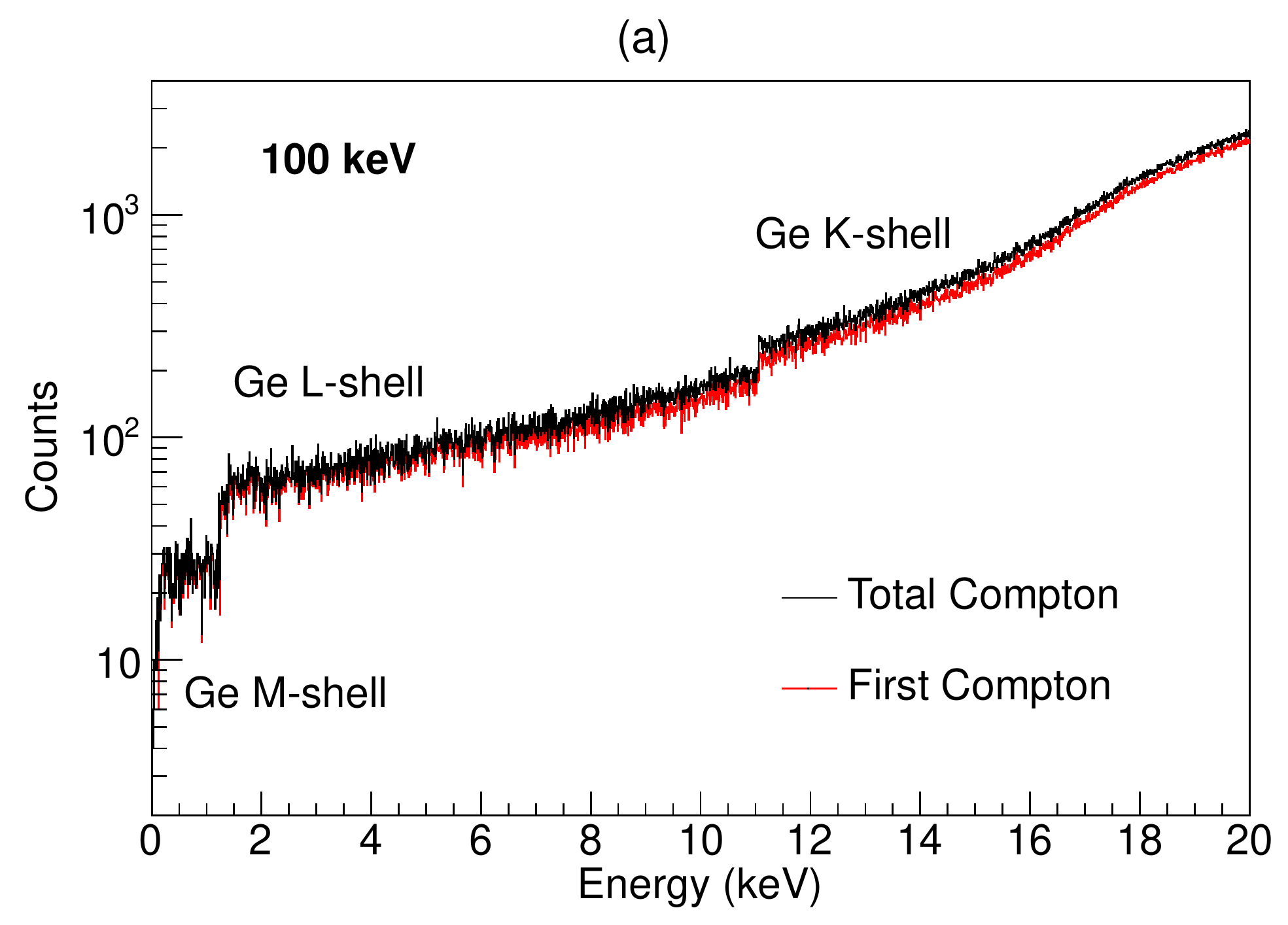}
\centering\includegraphics[width=\columnwidth]{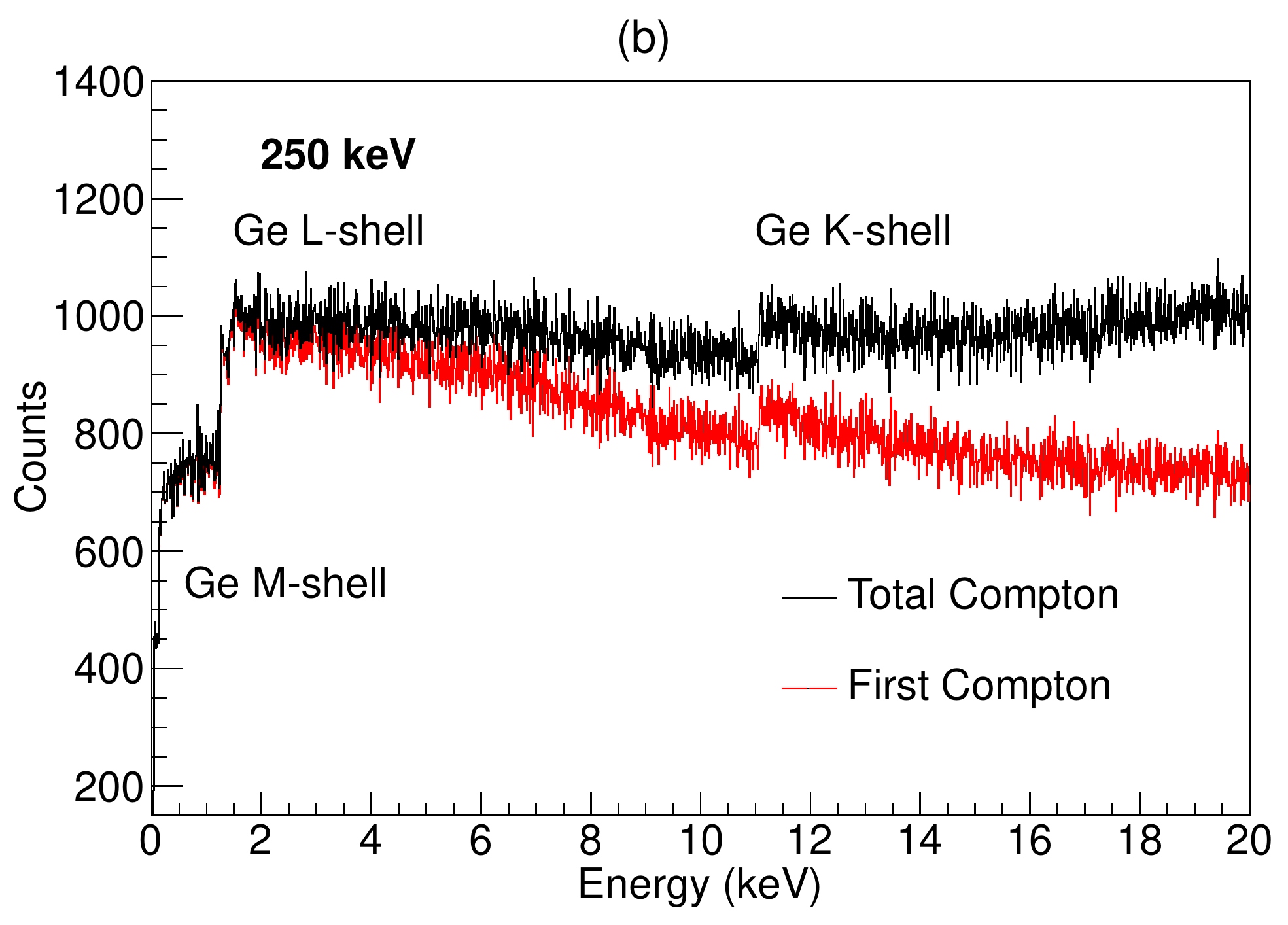}
\centering\includegraphics[width=\columnwidth]{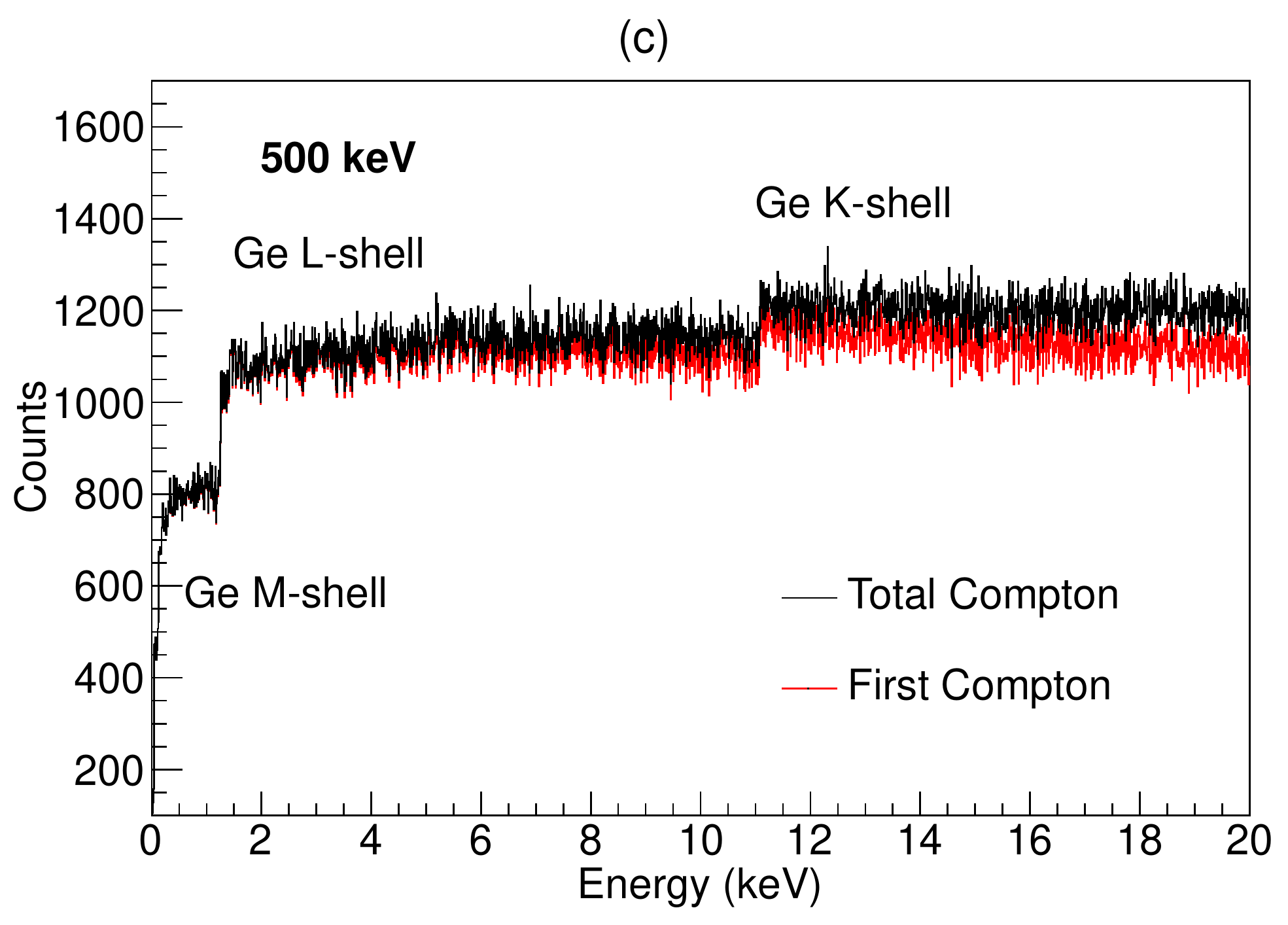}
\centering\includegraphics[width=\columnwidth]{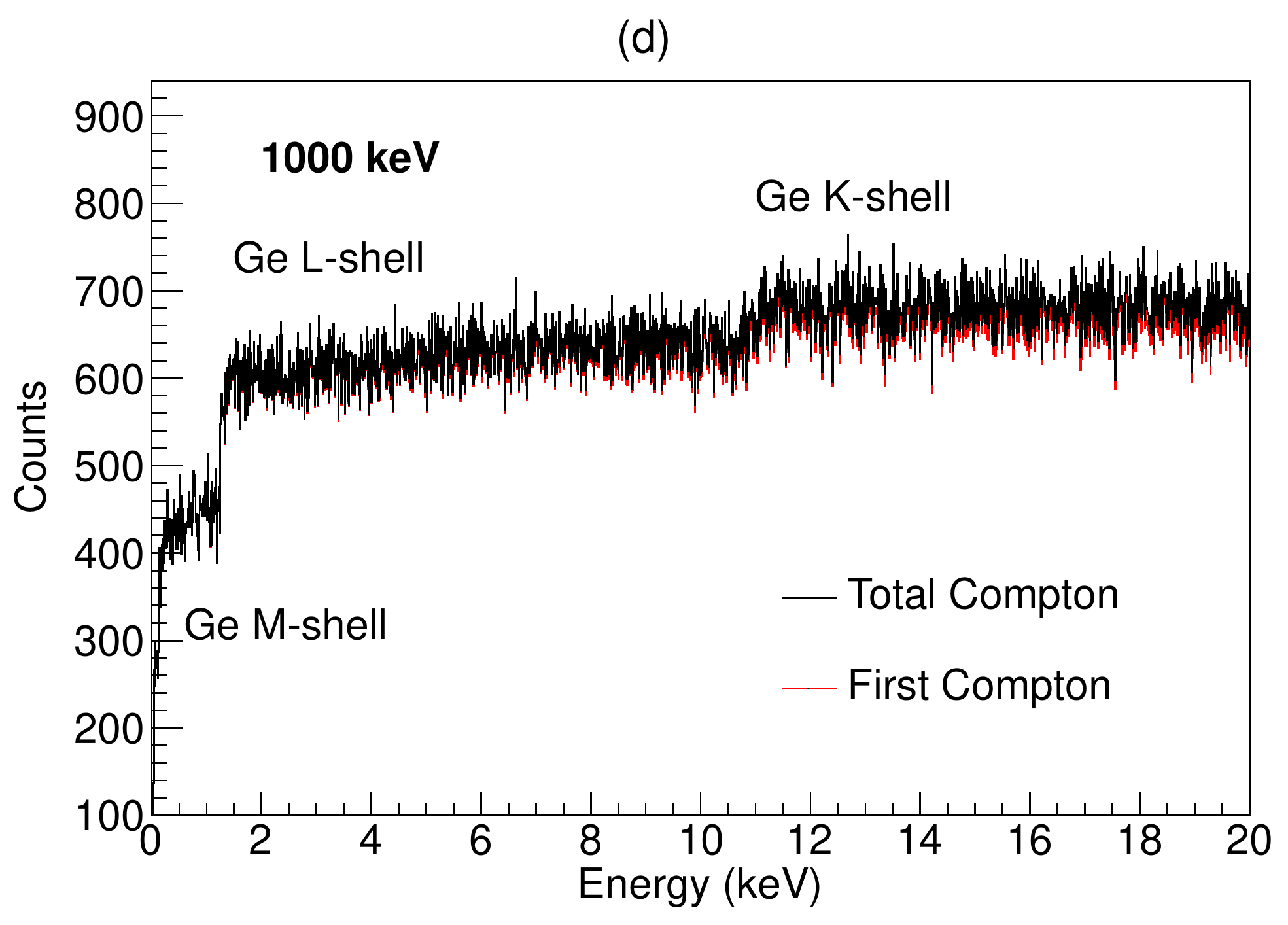}
\caption{Energy spectra at the low-energy region (0$–$20 keV) for different energies' $\gamma$ hitting the germanium crystal are shown in order: (a) 100 keV, (b) 250 keV, (c) 500 keV, and (d) 1000 keV. The black spectrum is the total Compton scattering deposition spectrum, and the red spectrum is the deposition spectrum of only one single Compton scattering.}
\label{fig:4}
\end{figure*}

It is undeniable that the Geant4 simulation (and thus RIA model) has a systematically sharper slope compared to the measured data and fitting. This may be caused by the known problem that the RIA model does not do well near the shells. More research is ongoing and detail explanations is postponed to future work.

\section{Discussion}\label{sec:5}
For low energy $\gamma$ rays, the Compton edge is low and the energy spectra at low energy will not flat, which may blur the K-shell step structure. It is necessary to study that how multi-scatters effect the spectrum. We performed more simulations to compare the energy spectra in the low-energy region induced by different gamma sources. Figure~\ref{fig:4} shows the energy spectra induced by the incidents 100 keV, 250 keV, 500 keV, and 1000 keV to the germanium crystal, respectively. The particle position is set 5 cm above the germanium crystal only, and the particle incidence direction is perpendicular to the crystal surface.  
There are large differences in the shape of the energy spectra for different energies of the incident $\gamma$ rays. To get the slope tendency after the K-shell, we performed a simulation using a single germanium crystal, where $\gamma$ rays were input from 100 keV to 3000 keV, and the Compton step slope of the K-shell of the total events and single Compton events was calculated based on different energy values, respectively. The calculation method was to fit the interval from 11.1--18 keV to obtain the K-shell slope of the three cases, respectively.

\begin{figure}[!htbp]
\centering\includegraphics[width=\columnwidth]{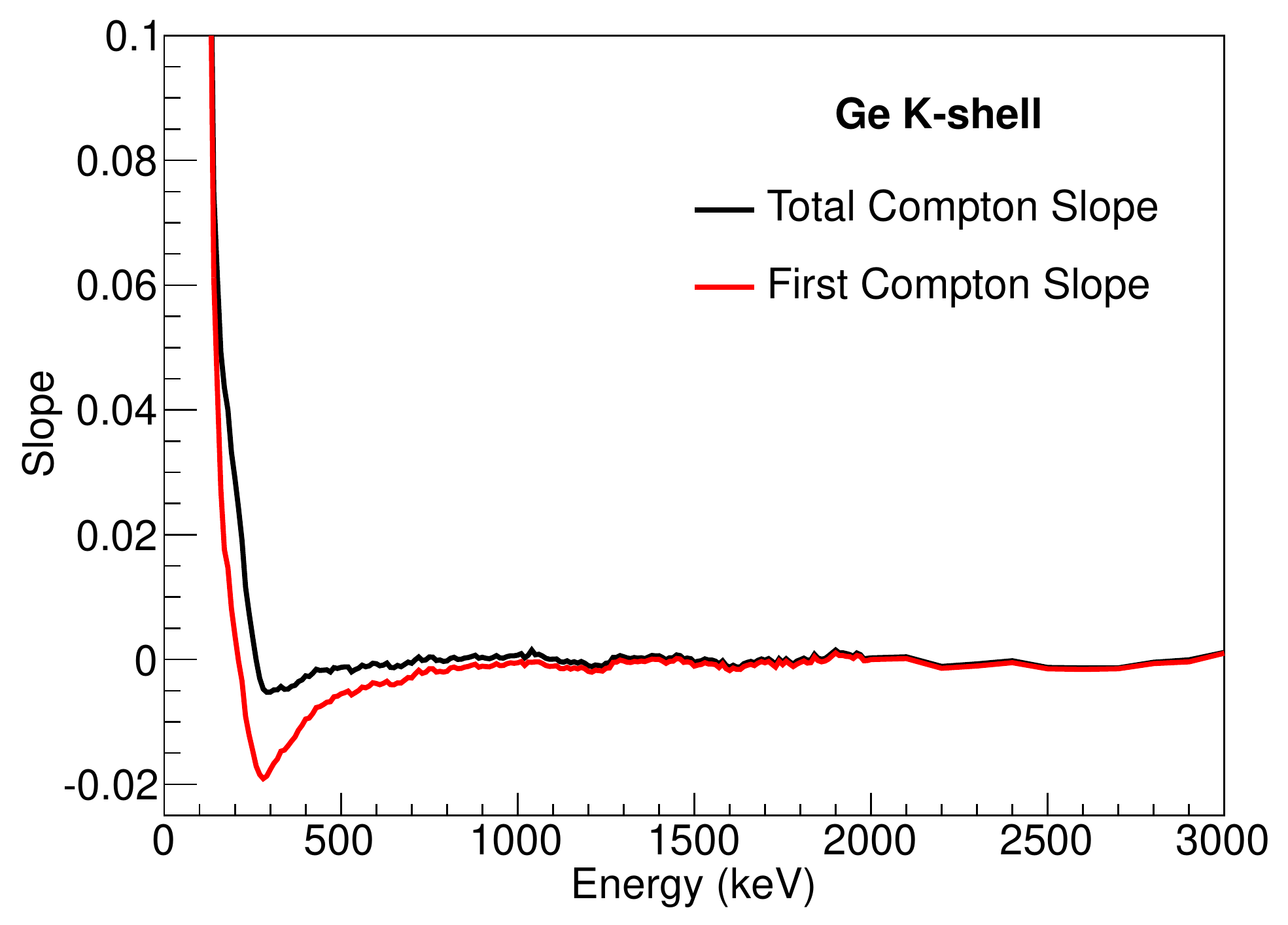}
\caption{Results of the Compton step slope of the K-shell from 100 keV to 3000 keV $\gamma$ for the total event and the single Compton scattering event.}
\label{fig:5}
\end{figure}

Figure~\ref{fig:5} shows the slopes of the total Compton event spectra decrease rapidly to 0 before 250 keV and fluctuate around 0 from 250 keV to 3000 keV. For the Compton-only event spectra, the slope values decreased more rapidly compared with the total Compton event spectra before 280 keV; however, they gradually increased after 280 keV. The increasing speed was proportional to the square of the incident $\gamma$ energy and reached $\sim$0 at last according to the RIA theory. It can be seen that the change trend of the slopes of the total Compton event was not exactly the same as that of the single Compton event. Below 150 keV, the slopes of the total Compton event spectra were the same as those of the Compton-only event spectra. In the vicinity of 150--1000 keV, the slopes of the total Compton event spectra and Compton-only event spectra were quite different, mainly due to the increase in the secondary Compton event, resulting in a difference in the slopes of the two. In the energy range of 1000--3000 keV, after the first Compton process with the germanium atoms, the secondary gammas still carried high enough energy to escape from the germanium crystal.

\section{Summary and prospects}\label{sec:6}
In this work, we obtained Compton scattering data in the low-energy region below 18 keV by irradiating the CDEX-1B pPCGe detector with two gamma sources, $^{137}$Cs and $^{60}$Co. Moreover, we observed the Compton step structure of the L-shell layer of the germanium atoms for the first time and the Compton step structure of both the L-shell and K-shell of the germanium atoms. According to experimental and simulation studies, the heights of the Compton steps were mainly related to the number of electrons in the different shell layers. The experimental data showed to be in good agreement with the simulated results.

In the analysis, we adopted a bin width of 200 eV and an analysis threshold of 500 eV, based on the principle of making the statistics in each bin as significant as possible and an acceptable systematic uncertainty during the data analysis procedure~\cite{cdex1b2018}, especially the bulk/surface discrimination~\cite{Yang:2018a}, which limits the bin width and analysis threshold under the exposure of these dataset in this study. Currently we are developing an unbinned bulk/surface discrimination method that has the potential to free us from the limit of bin width and statistical concerns, which is postponed to future work. Simultaneously, it is necessary to further lower the detector threshold to measure the M-shell step at 180 eV and improve the Compton step model in the low energy region.

It may also be necessary to improve the Compton scattering model at low energy in Geant4 codes. In Geant4, the database on Compton profiles as well as scattering functions still uses the calculations made by F. Biggs as well as J.H. Hubbell $et\ al.$ in the 1970's~\cite{Hubbell1975AtomicFF, BIGGS1975201}. J.H. Hubbell $et\ al.$ used Waller-Hartree theory to calculate and used Hatree-Fork theory to obtain the state of the atom. These methods do not take into account relativistic effects. If relativistic effects and more electronic correlations are considered, a more accurate computational model will be obtained. Recently, some measurements show better agreement with predictions from $ab~initio$ calculations of Compton scattering with bound electrons~\cite{Qiao_2020,DNorcini2022}. We will also perform the ab initio calculations to quantify the impact on the CDEX low-mass sensitivity. 

In ultralow background dark matter detection experiments, due to the complex radioactive environment of the detector and the superposition of energy spectra from different energies, the background energy spectra could blur the stepped structure and make it become a ``flat" structure. Moreover, the stepped structure would be blurred by many different sources of backgrounds to be a flat structure in the low-energy region. That is one possible reason why we did not observe the Compton steps, but only observed a ``flat'' background in the experimental data of previous dark matter analysis~\cite{cdex102018,cdex12016,cdex12014}. The measured background energy spectrum is indeed composed of the energy deposition from these $\gamma$ rays with different energies, together with other backgrounds. At the current background level, it is hard to see the Compton steps in the dark matter analysis data, but these Compton steps should be considered when understanding the background contribution and building the background model.

In future CDEX experiments, we will further reduce the environmental background by reducing the amount of material and the radioactive background contribution of the structure material around the detector, and the Compton step may be observed on the background experimental data in the future. Precision measurements of the Compton step will help us to establish the background model of the low-energy region, improve the experimental sensitivity, and provide robust knowledge of backgrounds from Compton scattering. Our results can be directly applicable to background estimates for the CDEX experiment, as well as for other direct searches for dark matter that employ germanium detectors.

\begin{acknowledgments}
This work was supported by the National Key Research and Development Program of China (Grants No. 2017YFA0402200, No. 2022YFA1605000) and the National Natural Science Foundation of China (Grants No. 12175112, No. 12005111, No. 11975159, and No. 11725522).
\end{acknowledgments}

\bibliography{ComptonStep.bib}

\end{document}